\documentclass[11pt]{article}
\newcommand{\blind}{1}

\usepackage{float}
\usepackage{graphicx}
\usepackage{subcaption}
\usepackage[round, sort]{natbib}
\usepackage{float}
\usepackage{geometry}
\usepackage{tikz}
\usepackage[english]{babel}
\usepackage{longtable}
\usepackage{color}
\usepackage{amssymb, amsmath, amsthm}
\usepackage{multirow}
\usepackage[neverdecrease]{paralist}
\usepackage[titletoc,title]{appendix}
\usepackage{authblk}
\usepackage{setspace}
\usepackage{enumerate}
\usepackage{graphicx}
\usepackage{multirow}
\usepackage{mathtools}

\DeclareMathOperator{\expit}{expit}

\DeclareMathOperator{\median}{median}
\DeclareMathOperator{\var}{Var}

\newcommand{\indep}{\rotatebox[origin=c]{90}{$\models$}}

\RequirePackage[OT1]{fontenc}

\newtheorem{theorem}{Theorem}
\newtheorem{lemma}{Lemma}

\newcommand{\Pn}{\mathbb{P}_n}

\newcommand{\plugin}{\hat\theta_{\text{\tiny OD}}}
\newcommand{\aipw}{\hat\theta_{\text{\tiny AIPW}}}

\newcommand{\tmle}{\hat\theta_{\text{\tiny TMLE}}}
\newcommand{\ipw}{\hat\theta_{\text{\tiny IPW}}}
\newcommand{\firpo}{\hat\theta_{\text{\tiny FIRPO}}}
\usepackage{hyperref}
\hypersetup{
  colorlinks=true,
  linkcolor=blue,
  filecolor=magenta,
  urlcolor=blue,
  citecolor=blue,
}
\title{Efficient Estimation of Quantiles in Missing Data Models}

\if1\blind \author[1,2]{Iv\'an D\'iaz \thanks{This research was
    completed during the author's tenure at Google Inc.; the
    manuscript was updated and revised after he joined Weill Cornell
    Medicine.}}  \affil[1]{\small Division of Biostatistics and
  Epidemiology, Weill Cornell Medicine, New York, NY, USA.}
\affil[2]{\small Google Inc., New York, NY, USA.}  \if0\blind
\author[1]{\vspace{-2cm} } \fi

\begin{document}\maketitle
\maketitle
\begin{abstract}
  We propose a novel targeted maximum likelihood estimator (TMLE) for
  quantiles in semiparametric missing data models. Our proposed
  estimator is locally efficient, $\sqrt{n}$-consistent,
  asymptotically normal, and doubly robust, under regularity
  conditions. We use Monte Carlo simulation to compare our proposed
  method to existing estimators. The TMLE is superior to all
  competitors, with relative efficiency up to three times smaller than
  the inverse probability weighted estimator (IPW), and up to two
  times smaller than the augmented IPW. This research is motivated by
  a causal inference research question with highly variable treatment
  assignment probabilities, and a heavy tailed, highly variable
  outcome. Estimation of causal effects on the mean is a hard problem
  in such scenarios because the information bound is generally
  small. In our application, the efficiency bound for estimating the
  effect on the mean is possibly infinite. This rules out
  $\sqrt{n}$-consistent inference and reduces the power for testing
  hypothesis of no treatment effect on the mean. In our simulations,
  using the effect on the median allows us to test a location-shift
  hypothesis with 30\% more power. This allows us to make claims about
  the effectiveness of treatment that would have hard to make for the
  effect on the mean. We provide R code to implement the proposed
  estimators.
\end{abstract}

\vspace{9pt}
\noindent {\it Key words:}
Quantile effects, information bound, $\sqrt{n}$-consistency, TMLE.
\par

\section{Introduction}

Estimation of quantiles in missing data models is a statistical
problem with applications to a variety of research areas, but has been
somewhat overlooked in the semiparametric inference literature. For
example, policy makers may be interested in evaluating the effect of
an educational program on the tails of the skill distribution. In this
case quantile treatment effects may be useful since they capture
intervention effects that are heterogeneous across the outcome
distribution. Quantiles may also be useful in economics research to
compute inequality indicators such as the Gini coefficient. Quantile
treatment effects may also be useful in the study of treatment effect
heterogeneity, e.g., for assessing whether all quantiles are equally
affected by treatment.

Our methods are motivated by an application to estimation of the
causal effect of treatment on an outcome whose distribution exhibits
heavy tails. The data we consider arises as part of various sales and
services programs targeted to introduce new features to users of the
AdWords advertisement platform at Google Inc. A important question for
decision makers is to quantify the causal effect of these programs on
the advertisers' spend through AdWords. The outcome we consider
exhibits heavy tails, as there is a small but non-trivial number of
advertisers who spend large quantities through AdWords. Heavy tailed
distributions are often characterized by large or infinite variance,
which in turn yields a large or infinite efficiency bound for
estimating the effect of treatment on the mean. As a consequence, the
variance of all regular estimators is also large, possibly precluding
$\sqrt{n}$-consistent inference and statistical significance at most
plausible sample sizes. Therefore, though the effect on the mean is
arguably an important parameter fro this problem (the mean spend is
directly related to total spend), $\sqrt{n}$-consistent inference for
it may be impossible or very hard in our application. In
Section~\ref{sec:aplica} we present simulation results showing that
the $n$-scaled mean squared error may not converge. This is a
strong indication of an infinite efficiency bound.

Estimation of quantiles in missing data models may be of general
interest in several scenarios, such as those described in the first
paragraph of this introduction, irrespective of the skewness of the
outcome. Our methods are motivated by using the effect on the
quantiles as an alternative to the effect on the mean, as a test
statistic for a location-shift hypothesis. We propose a novel targeted
maximum likelihood estimator. This estimator is locally efficient in
the non-parametric model and asymptotically linear, under certain
regularity conditions.  In our application, estimating a collection of
quantiles of interest (e.g., 25\%, 50\% and 75\%) allows us to make
statements about treatment effects, even though we would have
difficulty making similar statements for the mean, due to the large
variability caused by the heavy tailed distribution.

Our goal is to estimate an unconditional quantile. An alternative goal
is to estimate an outcome quantile conditional on the values of
certain covariates. Though we do not estimate conditional quantiles,
we use covariate information in order to correctly identify the
unconditional quantiles under the missing at random
assumption. Estimation of conditional quantiles is discussed, for
example, by \cite{koenker2005quantile, buchinsky1998recent,
  yu1998local}, among many others. Adaptation of our methods to
estimate quantiles within strata of categorical baseline covariates
are possible but are not discussed here.

In order to assess the performance of the proposed estimators in our
real data application, we use Monte Carlo simulations based on a real
dataset to approximate the bias, variance, mean squared error, and
coverage probability of the confidence interval estimators for our
application. Our proposed TMLE has the best performance across various
modeling scenarios in comparison to the available alternative of IPW
and augmented IPW (AIPW) estimation. We also use the simulation study to
demonstrate that estimation of the effect on the median has improved
power compared to the effect on the mean for testing the
location-shift hypothesis.

Our paper is organized as follows. In Section~\ref{sec:lit} we discuss
some existing approaches to estimation of quantiles in missing data
models. In Section~\ref{sec:notation} we formally introduce the
problem in terms of a closely related one: estimating the distribution
function of an outcome missing at random. In Section~\ref{sec:estima}
we present a summary of the available estimation methods, and present
our proposed estimators for the quantiles of a variable missing at
random as well as the effect of treatment on the quantiles, together
with a theorem providing the conditions for asymptotic linearity and
efficiency. In Section~\ref{sec:aplica} we present two simulation
studies, one with a synthetic data generating mechanism, and one using
a real dataset from our motivating application. The Monte Carlo
simulation study based on a real dataset is used to illustrate the
performance of our estimator and show the benefits of using the median
as a location parameter for the counterfactual distribution in the
presence of heavy tails. Finally, in Section~\ref{sec:discuss}, we
discuss some concluding remarks.

\section{Related Work}\label{sec:lit}

Various methods exist that address the problem considered
here. \cite{wang2010empirical} consider pointwise estimation of the
distribution function using the augmented inverse probability weighted
estimator applied to an indicator function, where the missingness
probabilities and observed outcome distribution functions are
estimated via kernel regression. They propose to use the distribution
function to estimate the relevant quantiles using an outcome
distribution estimator (i.e., the inverse of the estimated distribution
function). Their approach suffers from various flaws stemming from the
fact that the estimated distribution function may be ill-defined:
direct inverse probability weighting may generate estimates outside
$[0,1]$, and pointwise estimation may yield a non-monotonic
function. In addition, their approach may not be used in high
dimensions since kernel estimators suffer from the curse of
dimensionality.

\cite{zhao2013robust} propose similar estimators for non-ignorable
missing data, under the assumption that the missingness mechanism is
linked to the outcome through a parametric model that can be estimated
from external data sources. \cite{liu2011distribution},
\cite{cheng1996kernel}, and \cite{hu2011dimension} consider estimators
that yield estimated distribution functions in the parameter space,
relying either on kernel estimators for the outcome distribution
function, or knowledge of the true missingness
probabilities. \cite{firpo2007efficient} proposes to estimate the
quantiles by minimizing an inverse probability weighted check loss
function. Their estimator achieves non-parametric consistency by means
of a propensity score estimated as a logistic power series whose
degree increases with sample size. \cite{melly2006estimation},
\cite{frolich2013unconditional}, and \cite{chernozhukov2013inference}
consider estimation of the quantiles under a linear parametric model
for the distribution and quantile functions,
respectively. Their methods lack the double robustness and efficiency
properties of our proposal.

\cite{zhang2012causal} propose a variety of methods, including an IPW
and an AIPW. The AIPW estimator is expected to have similar asymptotic
properties to our proposed TMLE, i.e., it is expected to be doubly
robust, efficient, and asymptotically linear, under regularity
conditions. In the context of estimation of the effect on the mean,
targeted maximum likelihood estimators have consistently shown better
performance than their AIPW counterparts for finite samples \cite[see
e.g.,][]{porter2011relative}. In Section~\ref{sec:aplica} we show, in
simulation studies, that our proposed TMLE for the effect on the
quantiles also has superior finite sample performance.

\section{Notation and Estimation Problem}\label{sec:notation}
Let $Y$ denote an outcome observed only when a missingness indicator
$M$ equals one, and let $X$ denote a set of observed covariates
satisfying $Y\indep M\mid X$. We use $P_0$ to denote the true joint
distribution of the observed data $Z=(X, M, MY)$. Assume we observe an
i.i.d. sample $Z_1,\ldots, Z_n$, and denote its empirical distribution
by $\Pn$. We use the word \textit{model} to refer to a set of
probability distributions, and the expression \textit{nonparametric
  model} to refer to the set of all distributions having a continuous
density with respect to a dominating measure of interest. The word
\textit{estimator} is used to refer to a particular procedure or
method for obtaining estimates of $P_0$ or functionals of it. We
assume $P_0$ is in the nonparametric model $\cal M$, and use $P$ to
denote a general element of $\cal M$. For a function $h(z)$, we denote
$Ph=\int h dP$. For simplicity in the presentation we assume that $X$
is finitely supported but the results generalize to infinite support
by replacing the counting measure by an appropriate measure whenever
necessary. Under the assumption that $P_0(M=1\mid X=x) > 0$ almost
everywhere, the distribution $F_0(y)\equiv Pr(Y\leq y)$ is identified
in terms of $P_0$ as
\begin{align*} F_0(y)&=\sum_x Pr_0(Y\leq y\mid X=x)Pr_0(X=x)\\ &=\sum_x
  Pr_0(Y\leq y\mid M=1,X=x)Pr_0(X=x)\\ &=\sum_x G_0(y\mid 1,x)p_{X,0}(x),
\end{align*}
where we have denoted $G(y\mid 1,x)\equiv Pr(Y\leq y\mid M=1,X=x)$ and
$p_X(x)\equiv Pr(X=x)$. We use $f$ to denote the density corresponding
to $F$ and $e(x)$ to denote $Pr(M=1\mid X=x)$, following the
convention in the propensity score literature. We also denote
$\eta=(G,e)$. Consider the $q$-th quantile of the outcome
distribution:
\[\theta=\inf\{y: F(y)\geq q\}.\]
We use the notation $\theta(P)$ to
refer to the functional that maps an observed data distribution $P$
into a real number. Given a consistent estimator $\hat G$ of $G_0$,
the outcome distribution estimator $\plugin$ is obtained as an
(approximate) solution to the equation
$\frac{1}{n}\sum_{i=1}^n \hat G(\theta\mid, 1, X_i)=q$ is typically
consistent, but it may not be $\sqrt{n}$-consistent. Various methods
exist in the semi-parametric statistics literature that may be used to
remedy this issue. The analysis of the asymptotic properties of such
methods often relies on so-called von Mises expansions
\citep{mises1947asymptotic} and on the theory of asymptotic lower
bounds for estimation of regular parameters in semiparametric models
\cite[see, e.g.,][]{newey1990semiparametric, Bickel97}.

The efficient influence function $D(Z)$ is one of the key concepts introduced by
semi-parametric efficient estimation theory. This function characterizes all
efficient, asymptotically linear estimators $\hat\theta$. Specifically, the
following holds for any such estimator \citep[see e.g., ][]{Bickel97}:
\begin{equation} \sqrt{n}(\hat\theta - \theta) = \frac{1}{\sqrt{n}}\sum_{i=1}^n
  D(Z_i) + o_P(1/\sqrt{n}).\label{aslin}
\end{equation}
Asymptotic linearity allows the use of the central limit theorem to
construct Wald-type asymptotically valid confidence intervals and hypothesis
tests. For our target of inference $\theta$, the efficient influence
function in the non-parametric model is given below in
Lemma~\ref{lemma:eif}.
\begin{lemma}[Efficient Influence Function]\label{lemma:eif}
  The efficient influence function of $\theta$ at $P$ in the non-parametric model
  is equal to
  \begin{equation}\label{defD}
    D(Z)=-\frac{1}{f(\theta)}\left[\frac{M}{e(X)}\left\{I_{(-\infty, \theta]}(Y) -
        G(\theta\mid 1,X)\right\} + G(\theta\mid 1,X) - q\right].
  \end{equation}
\end{lemma}
When necessary, we favor the notation $D_{\eta, \theta}$ to indicate the
dependence of $D$ on the nuisance parameter $\eta=(G,e)$ and the
quantile $\theta$. Lemma~\ref{lemma:eif} is a direct consequence of the
functional delta method applied to the non-parametric estimator of
$F_0(y)$, and the Hadamard derivative of the quantile functional given
in Lemma 21.4 of \cite{van2000asymptotic}. Note that if there is no
missingness, then $P(M=1)=1$, and $D(Z)$ reduces to
$D(Z)=-(I_{(-\infty, \theta]}(Y) - q)/f(\theta)$.  Then (\ref{aslin}) is
the standard asymptotic linearity result for the sample median
\cite[see, e.g., corollary 21.5 of][]{van2000asymptotic}.

\begin{lemma}[Double Robustness of $D_{\eta,\theta}$]\label{lemma:dr}
  Let $\eta = (G,e)$ with either $G=G_0$ or $e=e_0$. Then
  $P_0D_{\eta,\theta_0} = 0$.
\end{lemma}
In the above lemma we established the double robustness of the
efficient influence function. As a consequence of this lemma, under
standard conditions for the analysis of $M$-estimators \cite[e.g.,
Theorem 5.9 of][]{van2000asymptotic}, an estimator $\hat\theta$ that
satisfies $\Pn D_{\hat \eta, \hat \theta}=0$ is consistent if either
$\hat G$ or $\hat e$ is consistent, but not necessarily both. This
argument motivates the construction of an AIPW estimator as
the (approximate) solution to $\Pn D_{\hat \eta, \theta}=0$ in
$\theta$, for auxiliary estimators $\hat G$ and $\hat e$. This
estimator was originally proposed by
\cite{zhang2012causal}. Specifically, the estimator is defined as an approximate
solution to the equation
\begin{equation}
  \frac{1}{n}\sum_{i=1}^n\left[\frac{M_i}{\hat e(X_i)}\left\{I_{(-\infty, \theta]}(Y_i) -
      \hat G(\theta\mid 1,X_i)\right\} + \hat G(\theta\mid 1,X_i)\right]=q\label{eq:defaipw}
\end{equation}
in $\theta$. We denote this estimator with
$\aipw$. 
Similarly, \cite{zhang2012causal}
proposed to estimate $\theta_0$ as the value $\ipw$ that solves the equation
\[\sum_{i=1}^n\frac{M_i}{\hat e(X_i)}I_{(-\infty, \theta]}(Y_i)=q\]
in $\theta$. In our simulation study of Section~\ref{sec:aplica} we
also consider the inverse probability weighted estimator proposed by
\cite{firpo2007efficient}, defined as
\[\firpo=\arg\max_\theta \sum_{i=1}^n\frac{M_i}{\hat e(X_i)}(Y_i-\theta)(I_{(-\infty,
  \theta]}(Y_i)-q),\]
for an estimate $\hat e$ of $e_0$.

\section{Targeted Maximum Likelihood Estimator}\label{sec:estima}

As an alternative to the above estimators, in this
section we propose a method to construct an estimator $\tilde P$, and
estimate $\theta_0$ with the substitution estimator
$\tmle=\theta(\tilde P)$. Our proposal is such that the component
$\tilde \eta$ of $\tilde P$ satisfies
\begin{equation}\label{Deq}
  \Pn D_{\tilde\eta,\tmle}=o_P(1/\sqrt{n}).
\end{equation}
Using $M$-estimation and empirical process theory we derive the
conditions under which this estimator is consistent, efficient, and
asymptotically normal. We present the proposed estimation algorithm
along with theoretical results establishing its asymptotic
properties. In our simulation studies of Section~\ref{sec:aplica}, we
use synthetic and real datasets to illustrate the superior
finite-sample performance of our estimator in comparison to the above
competitors.

Targeted maximum likelihood estimation is a general estimation method
concerned with the construction of substitution estimators that solve
a target estimating equation. The construction of a TMLE is carried
out in three steps as follows. First, estimate the nuisance parameter
$\eta$ by means of standard (possibly data-adaptive) prediction
techniques. Second, propose a parametric fluctuation submodel to
iteratively tilt the initial estimators towards a solution of the
target estimating equation. This submodel is such that its score spans
the components of the estimating function; its parameter is
estimated through standard maximum likelihood techniques. Because
maximum likelihood estimates solve the score equation, it follows that
the desired estimating equation is solved. Since the TMLE increases
the likelihood of the initial estimators in the direction of the
efficient influence function, it has been conjectured that it has
improved finite sample properties, compared to the AIPW
estimator. Such improvements have been illustrated for causal effects
on the mean in simulation
studies by \cite{gruber2010targeted, porter2011relative,
  stitelman2012general}, among others, and are illustrated in this
article for causal effects on the quantiles.

When the efficient influence function estimating equation is used to
construct the TMLE, the resulting estimator enjoys the same asymptotic
properties as the standard AIPW (e.g., double robustness,
efficiency), under standard regularity conditions.

The reader interested in further discussion and other technical
details underlying the general TMLE methodology is referred to
\cite{van2006targeted} and \cite{van2011targeted}. We now proceed to
define the estimator for our problem. Consider the following iterative
procedure:

\begin{enumerate}
\item \textit{Initialize.} Obtain initial estimates $\hat e$ and $\hat G$ of $e_0$
  and $G_0$. We discuss possible options to estimate these quantities
  in Section~\ref{sec:iniest} below.
\item \textit{Compute $\hat \theta$.} For the current estimate $\hat G$,
  compute
  \[\hat F(y)=\frac{1}{n}\sum_{i=1}^n \hat G(y\mid 1,X_i),\] and $\hat\theta=
  \inf\{y: \hat F(y)\geq q\}$.
\item \textit{Update $\hat G$.} Let $\hat g$ denote the density associated
  to $\hat G$, and consider the exponential submodel
  \[\hat g_\epsilon(y\mid 1,x)=c(\epsilon, \hat g)\exp\{\epsilon
  H_{\hat\eta,\hat\theta}(z)\}\hat g(y\mid 1,x),\] where $c(\epsilon, \hat g)$ is a normalizing
  constant and
  \[H_{\hat\eta,\hat\theta}(z) = \frac{1}{\hat e(x)}\{I_{(-\infty, \hat\theta]}(y) -
  \hat G(\hat\theta \mid 1,x)\}\] is the score of the model. Estimate $\epsilon$ as
  \[\hat\epsilon = \arg\max_\epsilon \sum_{i=1}^n M_i\log
  \hat g_\epsilon(Y_i\mid 1,X_i).\]
  The updated estimator of $g_0$ is
  given by $\hat g_{\hat \epsilon}$.
\item \textit{Iterate.} Let $\hat g = \hat g_{\hat \epsilon}$ and iterate steps
  2-3 until convergence.
\end{enumerate}
The TMLE of $\theta_0$ is denoted by $\tmle$ and is defined as
$\hat\theta$ in the last iteration. We also use $\tilde P$ to denote
the estimate of $P_0$ obtained in the last iteration.

The optimization problem in step 3 above is convex in $\epsilon$, so that under
regularity conditions we expect the algorithm to converge to the
global optimum. In our simulation studies we found it practical to
stop once $|\hat \epsilon |< 10^{-4}\times n^{-0.6}$.  Note
that the MLE of $\epsilon$ in the TMLE algorithm satisfies
\begin{equation} \sum_{i=1}^n\frac{M_i}{\hat e(X_i)}\{I_{(-\infty,
    \tmle]}(Y_i) - \tilde G(\tmle\mid 1,X_i)\}=o_P(1/\sqrt{n}).\label{scoreEq}
\end{equation}
This, together with the stopping criteria and the definition of the
TMLE as
\[\frac{1}{n}\sum_{i=1}^n \tilde G(\tmle \mid 1,X_i)=q,\]
yields $\Pn D_{\tilde\eta, \tmle}=o_P(1/\sqrt{n})$.
The following theorem, proved in Appendix~\ref{app:proofs},
establishes the consistency, asymptotic normality, and efficiency of
the TMLE.

\begin{theorem}[Asymptotic Distribution of $\tmle$]\label{theo:aslin} Let $\tmle$ and
  $\tilde G$ denote the TMLE of $\theta_0$ and $G_0$ as defined
  above. Let $||f||^2=\int fdP_0$ denote the squared $L_2(P_0)$
  norm. Denote $h_{0,\theta}(x) = G_0(\theta\mid 1, x)$ and $\tilde
  h_{\theta}(x) = \tilde G(\theta\mid 1, x)$, and assume
  \begin{enumerate}[(i)]
  \item $\tilde \eta=(\tilde G, \hat e)$ converges to some
    $\eta_1=(G_1, e_1)$ in the sense that
    $||\tilde h_{\theta_0} - h_{1,\theta_0}||\,||\hat e - e_1||=o_P(1/\sqrt{n})$, with
    either $h_{1,\theta_0}=h_{0, \theta_0}$ or $e_1 = e_0$.
  \item $P_0 D_{\tilde\eta, \theta_0}$ is an asymptotically linear
    estimator of the map $\eta\to P_0 D_{\eta_1, \theta_0}$ at $\eta=\eta_1$.
  \item The class of functions $\{D_{\eta,\theta}:|\theta-\theta_0|
    <\delta, ||h_{\theta}-h_{1, \theta}||<\delta, ||e-e_1||<\delta\}$ is Donsker for some
    $\delta >0$ and such that $P_0(D_{\eta,\theta} -
    D_{\eta_1,\theta_0})^2\to 0$ as $(\eta,\theta)\to
    (\eta_1,\theta_0)$.
  \end{enumerate}
  Then $\sqrt{n}(\tmle-\theta_0)\to N(0, \sigma^2)$. If $(\tilde G,
  \hat e)=(G_0, e_0)$, then $\tmle$ is efficient with $\sigma^2=\var\{D_{\eta_0,\theta_0}(Z)\}$.
\end{theorem}

\subsection{Initial Estimators}\label{sec:iniest}
Assumption~(i) of Theorem~\ref{theo:aslin} requires that at least one
of $G_0$ or $e_0$ is consistently estimated at a fast enough
rate. When the number of covariates is large, the curse of
dimensionality precludes the use of non-parametric estimators for
these parameters. An common approach is to make use of parametric
working models to estimate the nuisance parameters. Unfortunately,
these parametric models often fail to describe the complex relations
arising in large dimensional data sets and may therefore invalidate
the conclusions of an otherwise well designed study
\citep{starmans2011models}. In these scenarios, we advocate for the
use flexible, data adaptive estimators to fit these quantities, so
that assumption~(i) remains plausible. One such approach is given by
Super Learning \citep{vanderLaan&Polley&Hubbard07}. Super Learning is
an ensemble learning algorithm that works in three steps as
follows. First, a library of candidate estimators is proposed. This
library usually contains many flexible estimation algorithms, and may
contain some less flexible algorithms often hypothesized by
subject-matter experts based on a-priori scientific knowledge. Second,
the data is randomly split in a number of validation and training
sets. Each algorithm is then trained in each training set, with its
predictive performance estimated using the validation set. Lastly, the
estimated predictive performance of the prediction algorithms is used
to estimate their weights in a weighted convex combination of
predictive algorithms. The super learner is defined as the resulting
convex combination of candidate algorithms.

The super learner algorithm, discussed by \cite{Polley2011} in the
context of regression, may be used to estimate the probabilities
$e_0$. Estimation of a conditional expectation is a problem
extensively addressed in the statistical learning literature and we
omit further discussion here. In contrast, data-adaptive estimation of
a conditional density is a problem that has enjoyed considerably less
attention. In Appendix~\ref{app:cd} we discuss a super learning method
to estimate the density $g_0$ of $Y$ conditional on $(M=1, X=x)$.

\subsection{Estimating the Causal Effect on the Treated}
In this subsection we discuss estimation of the causal effect of treatment on
an outcome quantile among the treated. Specifically, let $X$ denote a set of
pre-treatment variables, let $T$ denote a binary variable indicating the
treatment group, and let $Y$ denote the outcome of interest. 
We define the potential outcomes $Y_t:t\in\{0, 1\}$ as the outcome
that would have been observed if, contrary to the fact, $P(T=t)=1$. We
assume that (i) $T\indep Y_0\mid X$, and that (ii)
$e(x)=P(T=1\mid X=x) < 1$ almost everywhere. Assumption (i) is often
referred to as the \textit{no unmeasured confounders} or
\textit{ignorability} assumption, and states that all factors that are
simultaneous causes of $T$ and $Y$ must be measured. Assumption (ii)
is referred to as the \textit{positivity} assumption, and ensures that
all units have a non-zero chance of falling in the control arm $T=0$
so that there is enough experimentation.  Note that $Y_1=Y$ on the
event $T=1$, so that the $q$-th quantile of $Y_1$ among units with
$T=1$ may be optimally estimated by the sample quantile of $Y$ among
treated units. Thus, we focus our attention on estimation of the
quantile of $Y_0$ among units with $T=1$.  Let
$F(y)=P(Y_0\leq y\mid T=1)$ denote the distribution function of
$Y_0$ conditional on $T=1$, then our target estimand is given by
\[\theta = \inf\{y:F(y)\geq q\}.\]
Under assumptions (i) and (ii) above, the distribution function
$F$ identified as
\[F(y)=\sum_x P_{Y}(y\mid 0,x)p_{X}(x\mid 1),\]
where $G(y\mid 0,x)= Pr(Y\leq y\mid T=0,X=x)$ and $p_X(x\mid 1)= Pr(X=x\mid T=1)$. The
efficient influence function for estimation of $\theta$ in the
non-parametric model may be found using similar techniques as in the previous
section as
\begin{multline}\label{defD1}
  D(Z)=-\frac{1}{f(\theta)}\left[\frac{1-T}{E(T)}\frac{e(X)}{1-e(X)}\left\{I_{(-\infty,
        \theta]}(Y) -
      G(\theta\mid 0,X)\right\} \right.\\+ \left.\frac{T}{E(T)}\{G(\theta\mid 0,X) -
    q\}\right],
\end{multline}
where $f$ is the probability density function associated to $F$. 

The targeted maximum likelihood estimation algorithm involves the
following steps:
\begin{enumerate}
\item \textit{Initialize.} Obtain initial estimates $\hat e$ and $\hat G$ of $e_0$
  and $G_0$.
\item \textit{Compute $\hat\theta$.} For the current estimate $\hat G$,
  compute
  \[\hat F(y)=\frac{1}{\sum_iT_i}\sum_{i=1}^n T_i\hat G(y\mid 0,X_i),\] and $\hat\theta=\inf\{y:\hat F(y)\geq q\}$.
\item \textit{Update $\hat G$.} Let $\hat g$ denote the density
  associated to $\hat G$, and consider the exponential model
  $\hat g_\epsilon(y\mid 0,x)=c(\epsilon, \hat g)\exp\{\epsilon
  H_{\hat\eta,\hat\theta}(z)\}\hat g(y\mid 0,x),$ where
  $c(\epsilon, \hat g)$ is a normalizing constant and
  \[\hat H_{\hat\eta,\hat\theta}(z) = \frac{\hat e(X)}{1-\hat
    e(x)}\{I_{(-\infty, \hat \theta]}(y) -
  \hat G(\hat\theta\mid 0,x)\}\] is the score of the model. Estimate $\epsilon$ as
  $\hat\epsilon = \arg\max_\epsilon \sum_{i=1}^n (1-T_i)\log
  \hat g_\epsilon(Y_i\mid 0,X_i).$ The updated estimator of $g$ is given by $\hat g_{\hat \epsilon}(y\mid 0,x).$
\item \textit{Iterate.} Let $\hat g = \hat g_{\hat\epsilon}$ and iterate steps
  2-3 until convergence.
\end{enumerate}
The TMLE of $\theta_0$ is denoted by $\tmle$ and is defined as the
value of $\hat\theta$ in the last iteration. Arguing as in the proof
for Theorem~\ref{theo:aslin} we find that this TMLE is asymptotically
linear, doubly robust, and locally efficient, under regularity
conditions.

\section{Simulation and Case Studies}\label{sec:aplica}

\subsection{Synthetic Data Simulation}\label{sec:synt}
In an controversial paper, \cite{kang2007demystifying} conducted a set
of simulations to study the performance of doubly robust estimators
for a missing data problem under misspecification of the outcome and
treatment models. In particular, they focus on a situation where the
weights $M_i/\hat e(X_i)$ are highly variable; a situation in which
standard AIPW estimators generally have poor performance \cite[see
also][]{robins2007comment}. This simulation setup was further
considered by \cite{porter2011relative} with the objective of
assessing the performance of various targeted maximum likelihood
estimators.

In this section we revisit the simulation of
\cite{kang2007demystifying} with the modified objective of estimating
the causal effect of a treatment variable on the median of the
potential outcomes. The data is generated as follows. Let
$W_1,\ldots,W_4$ be independent normally distributed variables with
mean zero and variance one. The treatment variable $T$ is generated
from a Bernoulli distribution with probability equal to
$\expit(-W_1+ 0.5\times W_2 - 0.25\times W_3 - 0.1\times Z_4)$, where
$\expit(x)=\{1 + \exp(-x)\}^{-1}$. The outcome is generated as
$Y=210+27.4\times W_1 + 13.7\times W_2 + 13.7\times W_3+ 13.7\times
W_4 + N(0,1)$. From this, we can determine the effect of $T$ on the
median of $Y$ as $\median(Y_1)-\median(Y_0)=0$. In average, $50\%$ of
the units are treated, with treatment probabilities in the interval
$[0.01, 0.98]$.  Estimators of the propensity score $e_0$ are
therefore expected to lead to inverse probability weights with high
variability, a situation in which doubly robust estimators may perform
poorly. The researcher does not observe the covariates $W$, but only
the following transformations:
\begin{align*}
  X_1 &= \exp(W_1/2)\\
  X_2 &=W_2/\{1 + \exp(W_1)\} + 10\\
  X_3 &=(W_1\times W_3/25 + 0.6)^3\\
  X_4 &=(W_2 + W_4 + 20)^2.
\end{align*}
As it is bound to happen under standard practice using standard
parametric models, inconsistent estimation of the outcome and
treatment mechanisms occurs when the observed variables $X$ are used
to fit linear (logistic) regression estimators. Consistent estimation
is achieved when the transformations $W(X)$ implied by the previous
display are used instead. We consider four modeling scenarios: (a)
$\hat G$ and $\hat e$ are consistent; (b) $\hat G$ is consistent but
$\hat e$ is not; (c) $\hat e$ is consistent but $\hat G$ is not; and
(d) both $\hat G$ and $\hat e$ are inconsistent. We generate $1000$
datasets for each sample size $n\in\{200,500\}$, and compute the TMLE,
AIPW, IPW, Firpo, and OD estimators for each dataset. We then use the
average and standard deviation of the 1000 estimates to approximate
the bias and MSE of the estimators. In the simulation, $\hat G$ is a
normal distribution with the corresponding estimated conditional mean
and residual variance. The code used to carry out this simulation is
presented in Appendix~\ref{app:code}.
\begin{table}[ht]
  \centering
  \begin{tabular}{clrrr|rrr}
    \hline
    &  & \multicolumn{3}{c|}{$n = 100$} & \multicolumn{3}{c}{$n = 500$} \\
    Scenario & Estimator & $\sqrt{\mbox{MSE}}$ & Bias & SD & $\sqrt{\mbox{MSE}}$ & Bias & SD \\
    \hline
    \multirow{4}{*}{(a)} & TMLE  & 1.33  & -0.00 & 1.33  & 0.71  & 0.01  & 0.71  \\
    & AIPW  & 1.42  & 0.01  & 1.42  & 0.71  & -0.00 & 0.71  \\
    & IPW   & 9.59  & -0.13 & 9.59  & 3.58  & 0.02  & 3.58  \\
    & Firpo & 7.65  & -0.11 & 7.65  & 3.02  & 0.12  & 3.01  \\
    & OD    & 0.29  & 0.02  & 0.29  & 0.11  & -0.01 & 0.11  \\ \hline
    \multirow{4}{*}{(b)} & TMLE  & 1.33  & 0.01  & 1.33  & 0.70  & 0.00  & 0.70  \\
    & AIPW  & 1.42  & -0.00 & 1.42  & 0.70  & -0.00 & 0.70  \\
    & IPW   & 10.65 & -5.32 & 9.23  & 6.34  & -5.38 & 3.36  \\
    & Firpo & 12.56 & -2.64 & 12.28 & 14.92 & 0.96  & 14.89 \\
    & OD    & 0.29  & 0.02  & 0.29  & 0.11  & -0.01 & 0.11  \\ \hline
    \multirow{4}{*}{(c)} & TMLE  & 6.13  & -1.50 & 5.94  & 2.63  & -0.33 & 2.61  \\
    & AIPW  & 7.18  & -0.95 & 7.12  & 2.98  & -0.22 & 2.98  \\
    & IPW   & 9.59  & -0.13 & 9.59  & 3.58  & 0.02  & 3.58  \\
    & Firpo & 7.65  & -0.11 & 7.65  & 3.02  & 0.12  & 3.01  \\
    & OD    & 8.33  & -7.21 & 4.16  & 7.68  & -7.46 & 1.82  \\ \hline
    \multirow{4}{*}{(d)} & TMLE  & 8.40  & -5.06 & 6.71  & 5.37  & -4.44 & 3.01  \\
    & AIPW  & 8.92  & -5.04 & 7.36  & 5.54  & -4.72 & 2.89  \\
    & IPW   & 10.65 & -5.32 & 9.23  & 6.34  & -5.38 & 3.36  \\
    & Firpo & 12.56 & -2.64 & 12.28 & 14.92 & 0.96  & 14.89 \\
    & OD    & 8.33  & -7.21 & 4.16  & 7.68  & -7.46 & 1.82  \\ \hline
  \end{tabular}
  \caption{Kang and Schafer simulation results. MSE is the mean squared error and SD is the standard deviation.}
  \label{table:ks}
\end{table}

\paragraph{Results} The results of the simulation are presented in
Table~\ref{table:ks}. The TMLE is seen to have better performance in
terms of MSE than all competitors across scenarios. This illustrates
our conjectured improved finite sample performance. In scenario (a),
the TMLE and AIPW are expected to have the same asymptotic
distribution. This is corroborated in this simulation at sample size
$n=500$. In all other scenarios, the TMLE and AIPW may have different
asymptotic distributions. When the treatment mechanism is
misspecified, the bias of the IPW and Firpo's estimators does not
disappear as $n$ increases. In addition, these estimators have a much
larger variability than the doubly robust estimators TMLE and
AIPW. The OD estimator has a much better performance than all other
estimators in scenarios (a)-(b), when the outcome distribution is
correctly estimated. This is a well known fact stemming from the fact
that the variance of the MLE in a parametric model is smaller than the
non-parametric efficiency bound. The OD estimator lacks the double
robustness property and is therefore inconsistent in scenarios (c)-(d).

\subsection{Real Data Simulation}

In this section we illustrate the finite sample performance of the
estimators in our motivating example. We estimate the effect on the
quantiles among the treated, which is defined in the previous
section. We use data from one of the AdWords programs at Google to
create a data generating mechanism that mimics key features of our
motivating applications such as high-dimensionality and heavy-tailed
outcomes. We assess the performance of our estimators using these data
generating mechanisms, which gives us a better idea of the expected
performance in our real datasets, in comparison with synthetically
generated data.

In our motivating problem, treatment consists of proactive
consultations by sales representatives that help identify advertisers'
business goals and suggest changes to improve performance. Since
advertisers do not always adopt the proposed changes, a unit is
considered treated if it is offered and accepts treatment. As a
result, treatment is not randomized and we must use methods for
observational data to assess the effect of these programs. The
original dataset consists of 40,303 units, with 29,362 being
treated. To adjust for confounders of the relation between treatment
and spend through AdWords, we use 93 variables containing baseline
characteristics of the customer as well as activity on their AdWords
account.

The sample size used in the simulations is $n=5,000$. This is a
relatively small sample size compared to the number of
covariates. Though admittedly smaller than the typical sizes seen in
our applications, a sample size of $5,000$ serves the present purpose of
illustrating the finite sample performance of the estimators, while
allowing for a computationally feasible simulation.

\begin{figure*}
  \centering
  \begin{subfigure}[t]{0.48\textwidth}
    \centering
    \includegraphics[scale = 0.39]{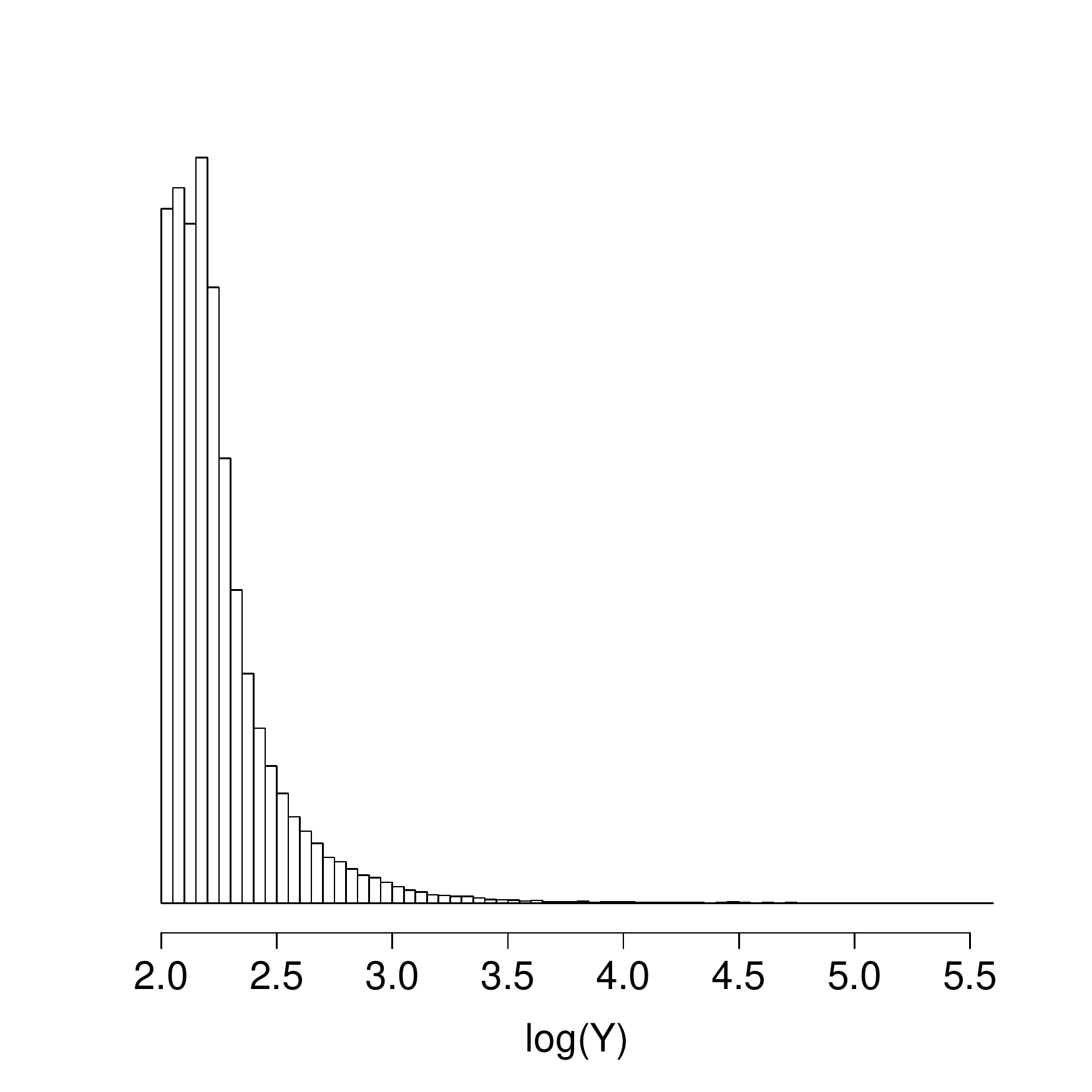}
    \caption{Histogram of the natural logarithm of the standardized outcome.}
    \label{hist}
  \end{subfigure}
  \begin{subfigure}[t]{0.48\textwidth}
    \centering
    \includegraphics[scale = 0.39]{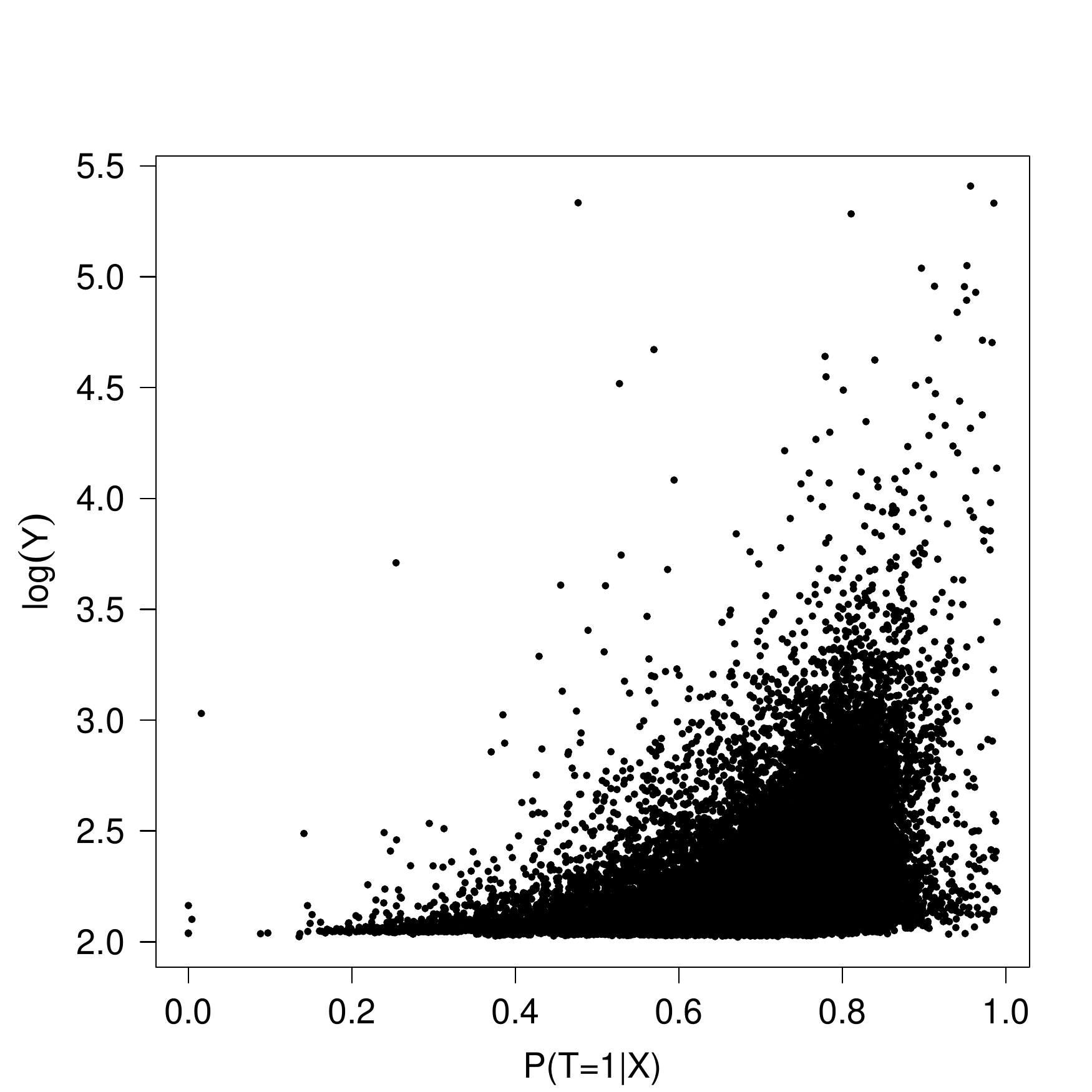}
    \caption{Log-outcome as a function of the propensity score.}
    \label{prop}
  \end{subfigure}
\end{figure*}

We have standardized the outcome to a variable with mean 10 and
standard deviation 5 before carrying out our analyses. These values
are selected arbitrarily and do not reflect any particular feature of
the data. Figure~\ref{hist} shows the distribution of the logarithm of
the standardized outcome, which can be seen to exhibit heavy tails and
a large variability, even in the logarithmic scale. In addition,
Figure~\ref{prop} shows that larger values of the outcome are
associated with larger propensities to receive treatment. This is
because, in our application, larger customers are more likely to
receive a sales consultation by a representative. Though it makes
practical sense from a business perspective, this poses an additional
challenge for estimation of causal effects since larger values of the
outcome are associated with smaller control probabilities. This
greatly increases the non-parametric efficiency bound for estimation
of the effect on the mean  \citep[see][for further discussion on this
issue]{robins2007}.

We use a re-sampling scheme based on parametric fits to the data in
order to recreate a scenario that closely resembles the real data
generating mechanism, while still allowing us to assess the
performance of the estimators under different types of model
misspecification with feasible computation times. We simulated 1,000
datasets from the observed data as follows. First, we fit a logistic
regression with main terms to the probability of treatment conditional
on the covariates on the real datasets. We then treat this logistic
regression function as the true probability of treatment conditional
on covariates. Second, we fit a main terms quantile regression to the
outcome, separately for the control and the treated group, for 500
quantiles, using the \texttt{quantreg} R package \citep{quantreg}.
The resulting quantile function is then treated as the true outcome
distribution conditional on treatment and covariates. We generate a
sample by first drawing covariates from the empirical distribution
(i.e., sampling with replacement). We then use the above probability
of treatment conditional on covariates to draw a treatment indicator,
and the above distribution of the outcome conditional on treatment and
covariates to draw an outcome value. We are interested in estimating
the effect of treatment on the 25\%, 50\%, and 75\% quantiles of the
outcome distribution. The true values for our data generating
mechanism are $0.10$, $0.13$, and $0.23$, respectively.

We compare the performance of the estimators in the same four
scenarios resulting from the correct or incorrect misspecification of
the estimators for $e_0$ and $G_0$ considered in
Section~\ref{sec:synt}. Misspecification of the estimators is carried out by
omitting 30 of the observed covariates when fitting the initial
estimators. The omitted covariates are chosen at random but fixed
through the simulation. Estimation of $\hat G$ is carried out by
fitting a parametric quantile regression algorithm on $500$ equally
spaced quantiles using the R package \texttt{quantreg}. As a result, the
initial density estimate $\hat g$ has point mass $1/500$ at each of
the initial quantiles, and the effect of the MLE in step 3 of
Section~\ref{sec:estima} is to update the probability mass of each
point. This algorithm is implemented in the R code presented in
Appendix~\ref{app:code}.

Estimator performance is assessed in terms of percent bias, variance,
mean squared error (MSE). For each generated dataset, we estimated the
effect of treatment on the $25\%$, $50\%$, and $75\%$ quantiles using
each of the estimators. We then approximate the bias, variance, and
MSE using empirical means across the 1,000 simulated datasets. The
results are presented in Table~\ref{res:simula}.

\paragraph{Results} The TMLE outperforms all its competitors in terms
of MSE, with the exception of the OD estimator when the outcome
distribution is consistently estimated. Though the OD estimator is
more efficient than the TMLE if the outcome distribution is estimated
consistently, it lacks the double robustness property. In general,
the OD estimator will only be $\sqrt{n}$-consistent in the unlikely case
that the outcome distribution is consistently estimated in a
parametric model.

We conjecture that the improved performance of the TMLE over the AIPW
is due to the property of the TMLE that it increases the likelihood of
the estimate $\hat G$ in the direction of the efficient influence
function, and therefore provides a better bias-variance trade-off. In
addition to having smaller variance, the TMLE has consistently a
smaller bias across simulation scenarios.

There are important efficiency gains obtained by using the TMLE in
comparison to its competitors. For example, for $q = 0.5$ and
scenario (i), the TMLE can deliver the same precision as $\firpo$
using $63\%$ fewer sample units. Similarly, the TMLE provides
important efficiency gains compared to the AIPW. For example, in
scenario (iii) and $q=0.75$, the TMLE attains a performance comparable
to the AIPW with $83\%$ fewer sample units.

\setlength{\tabcolsep}{4pt}
\begin{table}[!htb]
  \begin{tabular}{llrrr|rrr|rrr}
    \hline
    & & \multicolumn{3}{c|}{$q = 0.25$} & \multicolumn{3}{c|}{$q = 0.5$} & \multicolumn{3}{c}{$q = 0.75$} \\
    Scen. & Estim. & RMSE & Bias(\%)  & SD & RMSE & Bias(\%)  & SD & RMSE & Bias(\%) & SD  \\
    \hline
    \multirow{4}{*}{(a)} & TMLE  & 1.00 & 4.73   & 0.026 & 1.00 & 0.62   & 0.029 & 1.00 & -2.86  & 0.077 \\
    & AIPW  & 1.16 & 8.96   & 0.030 & 2.14 & 6.38   & 0.062 & 1.01 & 0.73   & 0.078 \\
    & IPW   & 1.95 & -2.82  & 0.052 & 3.16 & -5.71  & 0.092 & 1.60 & -17.35 & 0.117 \\
    & Firpo & 1.38 & -0.48  & 0.037 & 1.63 & -1.14  & 0.048 & 1.42 & 0.14   & 0.110 \\
    & OD    & 0.83 & -8.47  & 0.021 & 1.03 & -9.71  & 0.028 & 0.88 & -8.03  & 0.066 \\\hline
    \multirow{4}{*}{(b)} & TMLE  & 1.00 & 5.78   & 0.026 & 1.00 & 1.51   & 0.028 & 1.00 & -1.46  & 0.075 \\
    & AIPW  & 1.06 & 9.30   & 0.027 & 1.04 & 4.62   & 0.029 & 1.00 & 1.66   & 0.075 \\
    & IPW   & 1.65 & 27.05  & 0.036 & 1.77 & 17.54  & 0.045 & 1.46 & 11.51  & 0.106 \\
    & Firpo & 1.63 & 26.12  & 0.036 & 1.74 & 15.83  & 0.045 & 1.45 & 8.16   & 0.107 \\
    & OD    & 0.83 & -8.47  & 0.021 & 1.06 & -9.71  & 0.028 & 0.91 & -8.03  & 0.066 \\\hline
    \multirow{4}{*}{(c)} & TMLE  & 1.00 & 3.09   & 0.029 & 1.00 & -0.42  & 0.031 & 1.00 & -1.44  & 0.077 \\
    & AIPW  & 1.06 & 7.15   & 0.030 & 2.22 & 0.59   & 0.068 & 1.83 & 4.94   & 0.141 \\
    & IPW   & 1.82 & -2.82  & 0.052 & 2.99 & -5.71  & 0.092 & 1.60 & -17.35 & 0.117 \\
    & Firpo & 1.29 & -0.48  & 0.037 & 1.55 & -1.14  & 0.048 & 1.42 & 0.14   & 0.110 \\
    & OD    & 0.99 & -21.33 & 0.020 & 1.42 & -27.93 & 0.026 & 1.04 & -23.48 & 0.060 \\ \hline
    \multirow{4}{*}{(d)} & TMLE  & 1.00 & 10.06  & 0.028 & 1.00 & 3.90   & 0.029 & 1.00 & 1.19   & 0.074 \\
    & AIPW  & 1.08 & 14.24  & 0.029 & 1.07 & 7.75   & 0.030 & 1.04 & 5.26   & 0.075 \\
    & IPW   & 1.50 & 27.05  & 0.036 & 1.70 & 17.54  & 0.045 & 1.48 & 11.51  & 0.106 \\
    & Firpo & 1.49 & 26.12  & 0.036 & 1.67 & 15.83  & 0.045 & 1.47 & 8.16   & 0.107 \\
    & OD    & 0.97 & -21.33 & 0.020 & 1.47 & -27.93 & 0.026 & 1.09 & -23.48 & 0.060 \\ \hline
    \hline
  \end{tabular}
  \caption{Simulation results for different scenarios for the initial
    estimators (Scen.). \% Bias is the bias relative to the true
    parameter value. RMSE is the MSE relative to the MSE of the
    TMLE.}
  \label{res:simula}
\end{table}
\begin{table}[!htb]
  \begin{center}
    \begin{tabular}{c|rr|rr|rr}
      \hline
      & \multicolumn{2}{c|}{\% Bias} &
      \multicolumn{2}{c|}{$\sqrt{n\times \mbox{\small MSE}}$} &
      \multicolumn{2}{c}{Power}\\
      $n$    & Mean     & Median & Mean   & Median & Mean  & Median \\\hline

      5,000  & -3.541   & 2.178  & 5.202  & 2.621  & 0.649 & 0.944  \\
      10,000 & -4.082   & 0.787  & 4.750  & 2.605  & 0.894 & 0.999  \\
      20,000 & -4.283   & 0.683  & 5.302  & 2.387  & 0.982 & 1.000  \\
      40,000 & -4.566   & 0.490  & 5.843  & 2.355  & 0.995 & 1.000  \\
      \hline
    \end{tabular}
    \caption{Simulation results comparing TMLE of the effect on the mean
      vs the effect on the median as a measure of the causal effect of treatment
      among the treated.}
    \label{simula}
  \end{center}
\end{table}
\subsection{Testing The Location-shift Hypothesis}

In our motivating example, consider the location-shift hypothesis that
$p_{Y_1}(y\mid T=1) = p_{Y_0}(y - \beta \mid T=1)$, where $p_{Y_t}$
denotes the density function of the counterfactual variable $Y_t$. If
$\beta>0$, this hypothesis tells us that treatment had a positive
effect on the outcome by shifting the distribution of spend through
AdWords to the right. We are interested in comparing the performance of
estimators of the effect of the mean and the effect on the median as
test statistics for this hypothesis.  For the effect on the mean, we
focus on the TMLE for the average treatment effect on the treated
presented in Chapter 8 of \cite{van2011targeted}. This estimator
provides a fair competitor for our estimator of the effect on the
median among the treated, since it is also doubly robust and locally
efficient in the non-parametric model. 
A comparison of the relative MSE of the two estimators provides the
increase in sample size necessary to obtain comparable power.

We focus on a scenario with both models for $G_0$ and $e_0$  correctly specified. In this
scenario, in light of Theorem~\ref{theo:aslin}, we can use the
empirical variance
\scalebox{0.9}{$\widehat\var\{D_{\tilde \eta,\tmle}(Z)\}$} of the
estimated efficient influence function as a consistent estimator of
the TMLE. This, together with the asymptotic normality of $\tmle$,
allows us to perform Wald-type hypothesis tests of no treatment effect
on the median. A result analogous to Theorem~\ref{theo:aslin} is
available for the TMLE of the effect on the mean among the treated
\citep[see][for details]{van2011targeted}. Simulation scenarios with
misspecification of either model would require estimation of the
variance through the bootstrap and would imply prohibitive computation
times. Table~\ref{simula} contains a comparison between both
estimators in terms of their percent bias, the squared root of the
mean squared error scaled by $n$, and the power for testing the
hypothesis of no treatment effect.

Note the important loss of power for the test based on the mean as
compared to its median counterpart. The hypothesis test based on the
mean requires at least 2 times the sample size to achieve the power
obtained with the test based on the median. This is very relevant in
our setting since the sample size is not subject to modification but
rather fixed by the number of AdWords customers available in a certain
time period. In addition, the MSE of the estimator for the mean effect
scaled by $n$ seems to be increasing. Because the estimator used is
asymptotically efficient, a value of $n\times \mbox{MSE}$ that
diverges is a strong indication that the effect on the mean is not
estimable at a $\sqrt{n}$-rate.

\section{Concluding Remarks}\label{sec:discuss}

The TMLE algorithm proposed here is one of many possible ways to obtain a solution
to estimating equation (\ref{Deq}). Algorithms that aim at directly solving the
relevant estimating equation have been considered before. One option, discussed
by \cite{chaffee2011targeted} is to directly estimate the parameter $\epsilon$
as the solution to the estimating equation. A second option is to optimize the
log-likelihood function constrained to the set of parameters that solve
(\ref{Deq}). We favor the algorithm presented because, in addition to solving
the estimating equation, it guarantees an increase in the likelihood of the
final estimate $\tilde P$ compared to the initial $\hat P$. This has been shown to
yield estimators with improved finite sample properties
\citep{chaffee2011targeted}.

When the dimension of the baseline variables is large relative to the
sample size, the curse of dimensionality precludes the use of
nonparametric estimators for $e_0$ and $G_0$ \citep{Robins97R}. A
potential way to address this is to incorporate data-adaptive model
selection in constructing the initial estimators in step 1 of the TMLE
procedure, such as model stacking \citep{Wolpert1992} or super
learning \citep{vanderLaan&Polley&Hubbard07}. The asymptotic normality
of our estimator then require conditions (ii)-(iii) in
Theorem~\ref{theo:aslin}. These conditions would hold automatically
for the MLE in a parametric model, but need to be verified for
data-adaptive estimators. \cite{van2014targeted} proposed an estimator
for the case of a the mean in a missing data model that relaxes
assumption (ii); this approach is generalizable to our problem.

\begin{appendices}
\section{Proofs}\label{app:proofs}
\subsection{Lemma~\ref{lemma:dr}}
\begin{proof}
  By the law of iterated expectation we have
  \begin{align}
    P_0D_{\eta, \theta} &= -\frac{1}{f(\theta)}P_0\left[\frac{e_0}{e}(h_{0,\theta} -
      h_\theta) + h_\theta-q\right]\notag\\
    &= -\frac{1}{f(\theta)}P_0\left[\left(\frac{e_0}{e} - 1\right)(h_{0,\theta} -
      h_\theta) + (h_{0,\theta} - q)\right]\label{eq:R2},
  \end{align}
  where $h_{0,\theta}(x)=G_0(\theta\mid 1, x)$ and
  $h_{\theta}(x)=G(\theta\mid 1, x)$. The lemma follows from
  substituting either $G=G_0$ or $e=e_0$ in the previous display, with
  $\theta = \theta_0$.
\end{proof}

\subsection{Theorem~\ref{theo:aslin}}
\begin{proof}
  By construction of the TMLE algorithm we have
  $\Pn D_{\tilde \eta, \tmle} = o_P(1/\sqrt{n})$. By
  Lemma~\ref{lemma:dr} and Assumption (i) we have
  $P_0 D_{\eta_1, \theta_0} = 0$. In addition, by Theorem 5.9 of
  \cite{van2000asymptotic} we have $\tmle=\theta_0+o_P(1)$. Under
  assumptions (i)-(iii), an application of Theorem 5.31 of
  \cite{van2000asymptotic} yields
  \begin{equation}
    \sqrt{n}(\tmle-\theta_0)= \sqrt{n} P_0 D_{\tilde\eta, \theta_0} +
    \sqrt{n}(\Pn - P_0)D_{\eta_1, \theta_0} + o_P(1 + \sqrt{n}|P_0
    D_{\tilde \eta, \theta_0}|).\label{eq:wh}
  \end{equation}
  Let the influence function of Assumption~(ii) be denoted by
  $\Delta$. Then we have
  $\sqrt{n}P_0D_{\tilde\eta, \theta_0}= \sqrt{n}(\Pn - P_0)\Delta +
  o_P(1)$. Thus
  \begin{align*}
    \sqrt{n}(\tmle-\theta_0)&=
    \sqrt{n}(\Pn - P_0)(\Delta + D_{\eta_1, \theta_0}) + o_P(1 + O_P(1))\\
    &=
    \sqrt{n}(\Pn - P_0)(\Delta + D_{\eta_1, \theta_0}) + o_P(1).
  \end{align*}
  The central limit theorem yields the claimed asymptotic
  normality.

  Efficiency under the assumption that $\eta_1 = \eta_0$
  follows from the following argument. Equations~(\ref{eq:R2}) and the
  Cauchy-Schwartz inequality yields
  \[P_0D_{\tilde\eta, \theta_0} \leq C||\tilde h_{\theta_0} - h_{0, \theta_0}||\,||\hat e -
  e_0||,\]
  for some constant $C$. Under assumption (i) the term in the right hand side is
  $o_P(1/\sqrt{n})$. Then, from Equation~(\ref{eq:wh}) it follows that
  \begin{equation*}
    \sqrt{n}(\tmle-\theta_0)= \sqrt{n}(\Pn - P_0)D_{\eta_0, \theta_0} + o_P(1),
  \end{equation*}
  so that $\tmle$ is asymptotically normal, consistent, and efficient.
\end{proof}

\section{Super Learning for a Conditional Density}\label{app:cd}

The conditional density $g_0$ may be defined as
the minimizer of the negative log-likelihood loss function. That is
$g_0 = \arg\min_{f\in {\cal F}} R(f, p_0)$, where $\cal F$ is the
space of all non-negative functions of $(y,x)$ satisfying $\int
f(y,x)dy=1$, and $R(f)=-\int m\log f(y,x) dP_0(z)$. An estimator $\hat
g$ is seen here as an algorithm that takes a training sample ${\cal
  T}\subseteq \{Z_i:i=1,\ldots,n\}$ as an input, and outputs an estimated
function $\hat g(y\mid 1, x)$.

For a given estimator $\hat g$, we use cross-validation to construct
an estimate $\hat R(\hat g)$ of the risk $R(\hat g)$ as
follows. Let ${\cal V}_1,\ldots,{\cal V}_J$ denote a random partition
of the index set $\{1,\ldots,n\}$ into $J$ validation sets of
approximately the same size. That is,
${\cal V}_j\subset \{1,\ldots,n\}$;
$\bigcup_{j=1}^J {\cal V}_j = \{1,\ldots,n\}$; and
${\cal V}_j\cap {\cal V}_{j'}=\emptyset$. In addition, for each $j$,
the associated training sample is given by
${\cal T}_j=\{1,\ldots,n\}\setminus {\cal V}_j$. Denote by
$\hat g_{{\cal T}_j}$ the estimated density function obtained by
training the algorithm using only data in the sample ${\cal T}_j$. The
cross-validated risk of an estimated density $\hat g$ is defined as
\begin{equation}
  \hat R(\hat g) = -\frac{1}{J}\sum_{j=1}^J\frac{1}{|{\cal V}_j|}\sum_{i\in
    {\cal V}_j}\log M_i \hat g_{{\cal T}_j}(Y_i\mid 1, X_i).
\end{equation}
Consider now a finite collection ${\cal L}=\{\hat g_k:k=1,\ldots,K_n\}$ of
candidate estimators for $g_0$. We call this collection a {\em
  library}. We define the stacked predictor as a convex combination of
the predictors in the library:
\[\hat g_\alpha(y\mid 1, x)  =\sum_{k=1}^{K_n}\alpha_k\hat g_k(y\mid 1,
x), \]
and estimate the weights $\alpha$ as the minimizer of the cross-validated risk $\hat \alpha =
\arg\min \hat R(\hat g_\alpha)$, subject to
$\sum_{k=1}^{K_n}\alpha_k=1$. The final estimator is then defined as
$\hat g_{\hat\alpha}$.

\paragraph{Construction of the library}\label{sec:library}
Consider a partition of the range of $A$ into $k$ bins defined by a
sequence of values $\beta_0<\cdots<\beta_k$. Consider a candidate
for estimation of $g_0(y\mid 1, w)$ given by
\begin{equation}
  \hat g_\beta(y\mid 1, x)=\frac{\widehat{Pr}(Y \in
    [\beta_{t-1},\beta_{t})\mid M = 1, X=x)}{\beta_{t}-\beta_{t-1}},\text{ for } \beta_{t-1}\leq y <\beta_{t}.\label{eq:candidate}
\end{equation}
Here $\widehat Pr$ denotes an estimator of the true probability
$Pr_0(Y \in [\beta_{t-1},\beta_{t})\mid M=1, X=x)$ obtained through a
hazard specification and the use of an estimator for the expectation
of a binary variable in a repeated measures dataset as
follows. Consider the following factorization
\begin{multline*}
  Pr(Y \in [\beta_{t-1},\beta_{t})|M=1, X=x) = Pr(Y \in
  [\beta_{t-1},\beta_{t})|Y\geq \beta_{t-1},M=1,X=x)\times\\
  \prod_{j=1}^{t-1} \{1-Pr(Y \in [\beta_{j-1},\beta_{j})|Y\geq
  \beta_{j-1},M=1, X=x)\}.
\end{multline*}
The likelihood for model (\ref{eq:candidate}) is proportional to
\begin{multline*}
  \prod_{i=1}^nPr(Y_i \in [\beta_{t-1},\beta_{t})|M=1, X) =
  \prod_{i=1}^n\left[\prod_{j=1}^{t-1} \left\{1-Pr(Y_i \in [\beta_{j-1},\beta_{j})|Y_i\geq \beta_{j-1},M_i=1,X_i)\right\}\right]\times\\
  Pr(Y_i \in [\beta_{t-1},\beta_{t})|Y_i\geq \beta_{t-1},M_i=1,X_i),
\end{multline*}
which corresponds to the likelihood for the expectation of the binary
variable $I(Y_i \in [\beta_{j-1},\beta_{j}))$ in a repeated measures
data set in which the observation of subject $i$ is repeated $k$
times, conditional on the event $Y_i \geq \beta_{j-1}$.

Thus, each candidate estimator for $g_0$ is indexed by two choices:
the sequence of values $\beta_0<\cdots<\beta_k$, and the algorithm for
estimating the probabilities $Pr_0(Y_i \in
[\beta_{t-1},\beta_{t})|Y_i\geq \beta_{t-1},M_i=1,W_i)$. The latter is
simply a conditional probability, and therefore any standard
prediction algorithm may be used as a candidate. In the remainder of
this section we focus on the selection of the location and number of
bins, implied by the choice of $\beta_j$ values.

\cite{Denby&Mallows09} describe the histogram as a graphical
descriptive tool in which the location of the bins can be
characterized by considering a set of parallel lines cutting the graph
of the empirical cumulative distribution function
(ECDF). Specifically, given a number of bins $k$, the equal-area
histogram can be regarded as a tool in which the ECDF graph is cut by
$k+1$ equally spaced lines parallel to the $x$ axis. The usual
equal-bin-width histogram corresponds to drawing the same lines
parallel to the $y$ axis.  In both cases, the location of the cutoff
points for the bins is defined by the $x$ values of the points in
which the lines cut the ECDF. As pointed out by the authors, the
equal-area histogram is able to discover spikes in the density, but it
oversmooths in the tails and is not able to show individual
outliers. On the other hand, the equal-bin-width histogram oversmooths
in regions of high density and does not respond well to spikes in the
data, but is a very useful tool for identifying outliers and
describing the tails of the density.

As an alternative to find a compromise between these two approaches,
the authors propose a new histogram in which the ECDF is cut by lines
$x+cy=bh,\ b=1,\ldots,k+1$; where $c$ and $h$ are parameters defining
the slope and the distance between lines, respectively. The parameter
$h$ identifies the number of bins $k$. The authors note that $c=0$
gives the usual histogram, whereas $c\rightarrow \infty$ corresponds
to the equal-area histogram.

Thus, we can define a library of candidate estimators for the
conditional density in terms of (\ref{eq:candidate}) by defining values
of the vector $\beta$ through different choices of $c$ and $k$, and
considering a library for estimation of conditional
probabilities. Specifically, the library is given by the Cartesian
product
\[{\cal
  L}=\{c_1,\ldots,c_{m_c}\}\times\{k_1,\ldots,k_{m_k}\}\times\{\widehat
{Pr_1},\ldots, \widehat {Pr_{m_P}}\},\] where the first is a set of
$m_c$ candidate values for $c$, the second is a set of $m_k$ candidate
values for $k$, and the third is a set of $m_P$ candidates for the
probability estimation algorithm.  The use of this approach will
result in estimators that are able to identify regions of high density
as well as provide a good description of the tails and outliers of the
density. The inclusion of various probability estimators allows the
algorithm to find possible nonlinearities and higher order
interactions in the data. This proposed library may be augmented by
considering any other estimator. For example, there may be expert
knowledge leading to believe that a normal distribution (or any other
distribution) with linear conditional expectation could fit the
data. A candidate algorithm that estimates such a density using
maximum likelihood may be added to the library. This algorithm was
first proposed by \cite{Diaz11}, the reader interested in more details
and applications is encouraged to consult the original research
article.

\singlespacing
\section{R Code}\label{app:code}
{\small
\begin{verbatim}
trim <- function(x) pmax(x, 1e-10)

compute.quantile <- function(Q, w, q, r){
  F <- function(y)sapply(y, function(x)mean(rowSums((Q <= x) * w)))
  inv <- function(qq){
    uniroot(function(x){F(x) - qq}, r, extendInt = 'yes')$root
  }
  return(sapply(q, function(qq)inv(qq)))
}

od <- function(y, t, Q, g, q){
  n <- length(y)
  w  <- matrix(1/dim(Q)[2], ncol = dim(Q)[2], nrow = n)
  chiq <- compute.quantile(Q, w, q, range(y))
  return(chiq)

}

tmle <- function(y, t, Q, g, q){

  n <- length(y)
  D  <- function(y, w, chiq){
    1 / g * ((y <= chiq) - rowSums((Q <= chiq) * w))
  }
  w  <- matrix(1/dim(Q)[2], ncol = dim(Q)[2], nrow = n)
  h <- t
  chiq <- compute.quantile(Q, w, q, range(y))
  Do <- D(y, w, chiq)
  Dq <- D(Q, w, chiq)
  iter     <- 1
  crit     <- TRUE
  max.iter <- 20
  while(crit && iter <= max.iter){
    est.eq <- function(eps){
      out <- - mean(h * (Do - rowSums(Dq * exp(eps * Dq) * w) /
      rowSums(exp(eps * Dq) * w)))
      return(out)
    }
    loglik <- function(eps){
      out <- - mean(h * (eps * Do - log(rowSums(exp(eps * Dq) * w))))
      return(out)
    }
    eps <- optim(par = 0, loglik, gr = est.eq, method = 'BFGS')$par
    w <- exp(eps * Dq) * w / rowSums(exp(eps * Dq) * w)
    chiq <- compute.quantile(Q, w, q, range(y))
    Do <- D(y, w, chiq)
    Dq <- D(Q, w, chiq)
    iter <- iter + 1
    crit <- abs(eps)  > 1e-4 / n^0.6
  }
  return(chiq)
}

firpo <- function(y, t, Q, g, q){
  library(quantreg)
  h <- t / g
  chiq <- coef(rq(y ~ 1, weights = h, tau = q))
  names(chiq) <- NULL
  return(chiq)
}

ipw <- function(y, t, Q, g, q){
  n  <- length(y)
  w  <- matrix(1/dim(Q)[2], ncol = dim(Q)[2], nrow = n)
  h <- t / g
  D  <- Vectorize(function(chiq)  mean(h * (y <= chiq) - q))
  chiq <- uniroot(D, c(-1000, 1000), extendInt = 'yes')$root
  return(chiq)
}

aipw <- function(y, t, Q, g, q){
  n <- length(y)
  w <- matrix(1/dim(Q)[2], ncol = dim(Q)[2], nrow = n)
  h <- t / g
  D  <- Vectorize(function(chiq){
    mean(h * ((y <= chiq) - rowSums((Q <= chiq) * w))) +
    mean(rowSums((Q <= chiq) * w) - q)
  })
  chiq <- uniroot(D, c(-1000, 1000), extendInt = 'yes')$root
  return(chiq)
}

datagen <- function(n) {
  kBeta <- c(210, 27.4, 13.7, 13.7, 13.7)
  kTheta <- c(-1, 0.5, -0.25, -0.1)
  Z <- matrix(rnorm(n * 4), nrow = n, ncol = 4)
  X <- matrix(nrow = n, ncol = 4)
  X[, 1] <- exp(Z[, 1] / 2)
  X[, 2] <- Z[, 2] / (1 + exp(Z[, 1])) + 10
  X[, 3] <- (Z[, 1] * Z[, 3] / 25 + 0.6) ^ 3
  X[, 4] <- (Z[, 2] + Z[, 4] + 20) ^ 2
  y <- rnorm(n, mean = cbind(1, Z) %*% kBeta)
  true.prop <- 1 / (1 + exp(-Z %*% kTheta))
  T <- rbinom(n, 1, true.prop)
  dat <- list(Y = y, T = T, Z = Z, X = X)
  return(dat)
}

## Kang & Schafer Example

n.quant <- 500
formT <- T ~ X1+X2+X3+X4
formY <- Y ~ X1+X2+X3+X4
data <- datagen(1000)
X <- data$X
Y <- data$Y
T <- data$T
fitT <- glm(formT, data = data.frame(T=T, X), family = binomial)
fitY1 <- lm(formY, data = data.frame(Y=Y, T=T, X), subset = T == 1)
fitY0 <- lm(formY, data = data.frame(Y=Y, T=T, X), subset = T == 0)
median1 <- predict(fitY1, newdata = data.frame(T=1, X))
median0 <- predict(fitY0, newdata = data.frame(T=0, X))
Q1 <- sapply(seq(1/n.quant, 1 - 1/n.quant, 1/n.quant),
             function(q)qnorm(q, mean = median1, sd = summary(fitY1)$sigma))
Q0 <- sapply(seq(1/n.quant, 1 - 1/n.quant, 1/n.quant),
             function(q)qnorm(q, mean = median0, sd = summary(fitY0)$sigma))
g1 <- trim(predict(fitT, type = 'response'))
ame <- tmle(Y, T, Q1, g1, q) - tmle(Y, 1 - T, Q0, 1 - g1, q)


\end{verbatim}}
\end{appendices}
  \bibliographystyle{plainnat}
  \bibliography{bibliography}

\begin{thebibliography}{35}
\providecommand{\natexlab}[1]{#1}
\providecommand{\url}[1]{\texttt{#1}}
\expandafter\ifx\csname urlstyle\endcsname\relax
  \providecommand{\doi}[1]{doi: #1}\else
  \providecommand{\doi}{doi: \begingroup \urlstyle{rm}\Url}\fi

\bibitem[Bickel et~al.(1997)Bickel, Klaassen, Ritov, and Wellner]{Bickel97}
Peter~J. Bickel, Chris~A.J. Klaassen, Ya'acov Ritov, and Jon~A. Wellner.
\newblock \emph{Efficient and Adaptive Estimation for Semiparametric Models}.
\newblock Springer-Verlag, 1997.

\bibitem[Buchinsky(1998)]{buchinsky1998recent}
Moshe Buchinsky.
\newblock Recent advances in quantile regression models: a practical guideline
  for empirical research.
\newblock \emph{Journal of human resources}, pages 88--126, 1998.

\bibitem[Chaffee and van~der Laan(2011)]{chaffee2011targeted}
Paul Chaffee and Mark~J van~der Laan.
\newblock Targeted minimum loss based estimation based on directly solving the
  efficient influence curve equation.
\newblock 2011.

\bibitem[Cheng and Chu(1996)]{cheng1996kernel}
P.E. Cheng and C.K. Chu.
\newblock Kernel estimation of distribution functions and quantiles with
  missing data.
\newblock \emph{Statistica Sinica}, 6:\penalty0 63--78, 1996.

\bibitem[Chernozhukov et~al.(2013)Chernozhukov, Fern{\'a}ndez-Val, and
  Melly]{chernozhukov2013inference}
Victor Chernozhukov, Iv{\'a}n Fern{\'a}ndez-Val, and Blaise Melly.
\newblock Inference on counterfactual distributions.
\newblock \emph{Econometrica}, 81\penalty0 (6):\penalty0 2205--2268, 2013.

\bibitem[Denby and Mallows(2009)]{Denby&Mallows09}
L.~Denby and C.~Mallows.
\newblock Variations on the histogram.
\newblock \emph{Journal of Computational and Graphical Statistics}, Vol. 18,
  Iss. 1:\penalty0 21--31, 2009.

\bibitem[D\'iaz and van~der Laan(2011)]{Diaz11}
Iv\'an D\'iaz and Mark van~der Laan.
\newblock Super learner based conditional density estimation with application
  to marginal structural models.
\newblock \emph{The International Journal of Biostatistics}, 7\penalty0
  (1):\penalty0 38, 2011.

\bibitem[Firpo(2007)]{firpo2007efficient}
Sergio Firpo.
\newblock Efficient semiparametric estimation of quantile treatment effects.
\newblock \emph{Econometrica}, pages 259--276, 2007.

\bibitem[Fr{\"o}lich and Melly(2013)]{frolich2013unconditional}
Markus Fr{\"o}lich and Blaise Melly.
\newblock Unconditional quantile treatment effects under endogeneity.
\newblock \emph{Journal of Business \& Economic Statistics}, 31\penalty0
  (3):\penalty0 346--357, 2013.

\bibitem[Gruber and van~der Laan(2010)]{gruber2010targeted}
Susan Gruber and Mark~J van~der Laan.
\newblock A targeted maximum likelihood estimator of a causal effect on a
  bounded continuous outcome.
\newblock \emph{The International Journal of Biostatistics}, 6\penalty0 (1),
  2010.

\bibitem[Hu et~al.(2011)Hu, Follmann, and Qin]{hu2011dimension}
Zonghui Hu, Dean~A Follmann, and Jing Qin.
\newblock Dimension reduced kernel estimation for distribution function with
  incomplete data.
\newblock \emph{Journal of statistical planning and inference}, 141\penalty0
  (9):\penalty0 3084--3093, 2011.

\bibitem[Kang and Schafer(2007)]{kang2007demystifying}
Joseph~DY Kang and Joseph~L Schafer.
\newblock Demystifying double robustness: A comparison of alternative
  strategies for estimating a population mean from incomplete data.
\newblock \emph{Statistical science}, pages 523--539, 2007.

\bibitem[Koenker(2005)]{koenker2005quantile}
Roger Koenker.
\newblock \emph{Quantile regression}.
\newblock Number~38. Cambridge university press, 2005.

\bibitem[Koenker(2013)]{quantreg}
Roger Koenker.
\newblock \emph{quantreg: Quantile Regression}, 2013.
\newblock URL \url{http://CRAN.R-project.org/package=quantreg}.
\newblock R package version 5.05.

\bibitem[Liu et~al.(2011)Liu, Liu, and Zhou]{liu2011distribution}
Xu~Liu, Peixin Liu, and Yong Zhou.
\newblock Distribution estimation with auxiliary information for missing data.
\newblock \emph{Journal of Statistical Planning and Inference}, 141\penalty0
  (2):\penalty0 711--724, 2011.

\bibitem[Melly(2006)]{melly2006estimation}
Blaise Melly.
\newblock Estimation of counterfactual distributions using quantile regression.
\newblock \emph{Review of Labor Economics}, 68\penalty0 (4):\penalty0 543--572,
  2006.

\bibitem[Newey(1990)]{newey1990semiparametric}
Whitney~K. Newey.
\newblock Semiparametric efficiency bounds.
\newblock \emph{Journal of applied econometrics}, 5\penalty0 (2):\penalty0
  99--135, 1990.

\bibitem[Polley et~al.(2011)Polley, Rose, and van~der Laan]{Polley2011}
Eric~C. Polley, Sherri Rose, and Mark~J. van~der Laan.
\newblock Super learning.
\newblock In Mark~J. van~der Laan and Sherri Rose, editors, \emph{Targeted
  Learning: Causal Inference for Observational and Experimental Data}, pages
  43--66. Springer New York, New York, NY, 2011.

\bibitem[Porter et~al.(2011)Porter, Gruber, van Der~Laan, and
  Sekhon]{porter2011relative}
Kristin~E Porter, Susan Gruber, Mark~J van Der~Laan, and Jasjeet~S Sekhon.
\newblock The relative performance of targeted maximum likelihood estimators.
\newblock \emph{The International Journal of Biostatistics}, 7\penalty0
  (1):\penalty0 1--34, 2011.

\bibitem[Robins et~al.(2007{\natexlab{a}})Robins, Sued, Lei-Gomez, and
  Rotnitzky]{robins2007}
James Robins, Mariela Sued, Quanhong Lei-Gomez, and Andrea Rotnitzky.
\newblock Comment: Performance of double-robust estimators when “inverse
  probability” weights are highly variable.
\newblock \emph{Statist. Sci.}, 22\penalty0 (4):\penalty0 544--559, 11
  2007{\natexlab{a}}.

\bibitem[Robins et~al.(2007{\natexlab{b}})Robins, Sued, Lei-Gomez, and
  Rotnitzky]{robins2007comment}
James Robins, Mariela Sued, Quanhong Lei-Gomez, and Andrea Rotnitzky.
\newblock Comment: Performance of double-robust estimators when" inverse
  probability" weights are highly variable.
\newblock \emph{Statistical Science}, 22\penalty0 (4):\penalty0 544--559,
  2007{\natexlab{b}}.

\bibitem[Robins and Ritov(1997)]{Robins97R}
James~M Robins and Ya'acov Ritov.
\newblock Toward a curse of dimensionality appropriate (coda) asymptotic theory
  for semi-parametric models.
\newblock \emph{Statistics in Medicine}, 16\penalty0 (3):\penalty0 285--319,
  1997.

\bibitem[Starmans(2011)]{starmans2011models}
Richard~JCM Starmans.
\newblock Models, inference, and truth: probabilistic reasoning in the
  information era.
\newblock In Mark van~der Laan and Sherri Rose, editors, \emph{Targeted
  Learning: Causal Inference for Observational and Experimental Data}.
  Springer, 2011.

\bibitem[Stitelman et~al.(2012)Stitelman, De~Gruttola, and van~der
  Laan]{stitelman2012general}
Ori~M Stitelman, Victor De~Gruttola, and Mark~J van~der Laan.
\newblock A general implementation of tmle for longitudinal data applied to
  causal inference in survival analysis.
\newblock \emph{The international journal of biostatistics}, 8\penalty0 (1),
  2012.

\bibitem[van~der Laan(2014)]{van2014targeted}
Mark~J van~der Laan.
\newblock Targeted estimation of nuisance parameters to obtain valid
  statistical inference.
\newblock \emph{The International Journal of Biostatistics}, 10\penalty0
  (1):\penalty0 29--57, 2014.

\bibitem[van~der Laan and Rose(2011)]{van2011targeted}
Mark~J. van~der Laan and Sherri Rose.
\newblock \emph{Targeted learning: causal inference for observational and
  experimental data}.
\newblock Springer Science \& Business Media, 2011.

\bibitem[van~der Laan and Rubin(2006)]{van2006targeted}
Mark~J. van~der Laan and Daniel Rubin.
\newblock Targeted maximum likelihood learning.
\newblock \emph{The International Journal of Biostatistics}, 2\penalty0 (1),
  2006.

\bibitem[van~der Laan et~al.(2007)van~der Laan, Polley, and
  Hubbard]{vanderLaan&Polley&Hubbard07}
Mark~J. van~der Laan, Eric Polley, and Alan Hubbard.
\newblock Super learner.
\newblock \emph{Statistical Applications in Genetics \& Molecular Biology},
  6\penalty0 (25), 2007.
\newblock ISSN 1.

\bibitem[van~der Vaart(2000)]{van2000asymptotic}
Aad~W. van~der Vaart.
\newblock \emph{Asymptotic statistics}, volume~3.
\newblock Cambridge university press, 2000.

\bibitem[von Mises(1947)]{mises1947asymptotic}
Richard von Mises.
\newblock On the asymptotic distribution of differentiable statistical
  functions.
\newblock \emph{The annals of mathematical statistics}, pages 309--348, 1947.

\bibitem[Wang and Qin(2010)]{wang2010empirical}
Qihua Wang and Yongsong Qin.
\newblock Empirical likelihood confidence bands for distribution functions with
  missing responses.
\newblock \emph{Journal of Statistical Planning and Inference}, 140\penalty0
  (9):\penalty0 2778--2789, 2010.

\bibitem[Wolpert(1992)]{Wolpert1992}
David~H Wolpert.
\newblock Stacked generalization.
\newblock \emph{Neural Networks}, 5\penalty0 (2):\penalty0 241--259, 1992.

\bibitem[Yu and Jones(1998)]{yu1998local}
Keming Yu and MC~Jones.
\newblock Local linear quantile regression.
\newblock \emph{Journal of the American statistical Association}, 93\penalty0
  (441):\penalty0 228--237, 1998.

\bibitem[Zhang et~al.(2012)Zhang, Chen, Troendle, and Zhang]{zhang2012causal}
Zhiwei Zhang, Zhen Chen, James~F Troendle, and Jun Zhang.
\newblock Causal inference on quantiles with an obstetric application.
\newblock \emph{Biometrics}, 68\penalty0 (3):\penalty0 697--706, 2012.

\bibitem[Zhao et~al.(2013)Zhao, Tang, and Tang]{zhao2013robust}
Pu-Ying Zhao, Man-Lai Tang, and Nian-Sheng Tang.
\newblock Robust estimation of distribution functions and quantiles with
  non-ignorable missing data.
\newblock \emph{Canadian Journal of Statistics}, 41\penalty0 (4):\penalty0
  575--595, 2013.

\end{thebibliography}

\end{document}